\documentclass[a4paper,11pt]{article}
\usepackage{siunitx}
\usepackage{pos}
\usepackage{bbm}
\usepackage{setspace}

\newcommand{\dppp}{D^+ \to  \pi^-\pi^+\pi^+}

\newcommand{\dkkp}{D^+ \to  K^-K^+\pi^+}
\newcommand{\dpp}{D^0 \to  \pi^+\pi^-}
\newcommand{\dkk}{D^0 \to  K^+K^-}

\title{Unitary $KK$ and $K\pi$ inputs for an amplitude analysis of the $\dkkp$ decay}

\author*[a]{Felipe Cesario Laterza Lopes}
\author[b]{Gustavo Alejandro Loachamín Ordoñez}%
\author[a]{Patrícia Camargo Magalhães}

\renewcommand{\printHeadAuthors}{Felipe C. L. Lopes}

\affiliation[a]{Instituto de Física Gleb Wataghin, Universidade Estadual de Campinas (UNICAMP),\\
  Rua Sérgio Buarque de Holanda 777, Campinas, Brazil}

\affiliation[b]{Departamento de Física, Pontifícia Universidade Católica do Rio de Janeiro (PUC-Rio),\\
Rua Marquês de São Vicente 225, Rio de Janeiro, Brazil}

\emailAdd{felipe.lopes@dac.unicamp.br}
\emailAdd{gustavo.loachamin@aluno.puc-rio.br}
\emailAdd{pcmaga@unicamp.br}

\abstract{Charge-Parity Violation (CPV) is expected in three-body charm decays. In particular, in the singly Cabibbo suppressed decay process $\dkkp$; in which an amplitude analysis is required to understand the dynamics of Final State Interactions (FSI) and the mechanism that can generate CPV. In this work, we show the prospects of analysing recent data from the Large Hadron Collider beauty (LHCb) experiment at CERN, with a precise theoretical model expected to reveal CPV mechanisms. Previous analyses of the $\dkkp$ decay were unable to satisfactorily describe the data using the Isobar Model. Currently, LHCb has a sample of charmed meson decays with around 100 million events for the $\dkkp$ decay, so a more robust phenomenological model is needed to describe such a large data set. The present model has allowed the inclusion of hadronic FSI and will, in prospect, introduce three-body effects, overcoming some limitations of the Isobar Model, and has all factors necessary to reveal the presence of CPV. To do so, it factorizes the weak vertex separated in leading topologies and relies on extensions of Resonant Chiral Perturbation Theory to describe FSI. As a proof-of-concept, a toy dataset of the S-wave of the most recent amplitude analysis of $\dkkp$, by the CLEO-c collaboration, was fitted with this model. 
From fits to model-independent parametrizations of scattering amplitudes, the parameters of the FSI amplitudes were fixed, and the production parameters of the weak vertex were then obtained from the fit to the toy dataset, yielding a good agreement between the data and our model.
}

\FullConference{43rd International Conference on High Energy Physics (ICHEP 2026)\\
30 July  to 5 August 2026\\
Natal, Brazil\\}

\begin{document}
\maketitle

\section{Introduction}
\vspace{-0.3cm}
The violation of Charge-Parity (CP) symmetry has been observed in many systems of decaying hadrons, including D-meson decays~\cite{LHCb_charmCPV} in 2019. This observation was made by the LHCb Collaboration at CERN, an experiment dedicated to investigating Charge-Parity Violation (CPV) in heavy meson decays. An amplitude analysis might shed some light on the possibility of CPV in intermediate states of the $\dkkp$ decay. 

In this work, a dynamic model with inputs from $KK$ and $K\pi$ unitary amplitudes is proposed to conduct an amplitude analysis of the $\dkkp$ decay using data collected by the Large Hadron Collider beauty (LHCb) Experiment during Run 2 of the Large Hadron Collider (LHC). For the proposal of such model, a brief overview of CPV in charm decays is done in Sec. \ref{sec:charmcpv}, a description of the formalism used for an amplitude analysis in Sec. \ref{sec:ampan}, and finally a description of the model and its proof-of-concept in Sec. \ref{sec:model}. The conclusions are given in Sec. \ref{sec:conclusion}.

\section{CP Violation in charm decays}
\label{sec:charmcpv}
\begin{figure}[ht]
    \centering
    \includegraphics[width=0.2\linewidth]{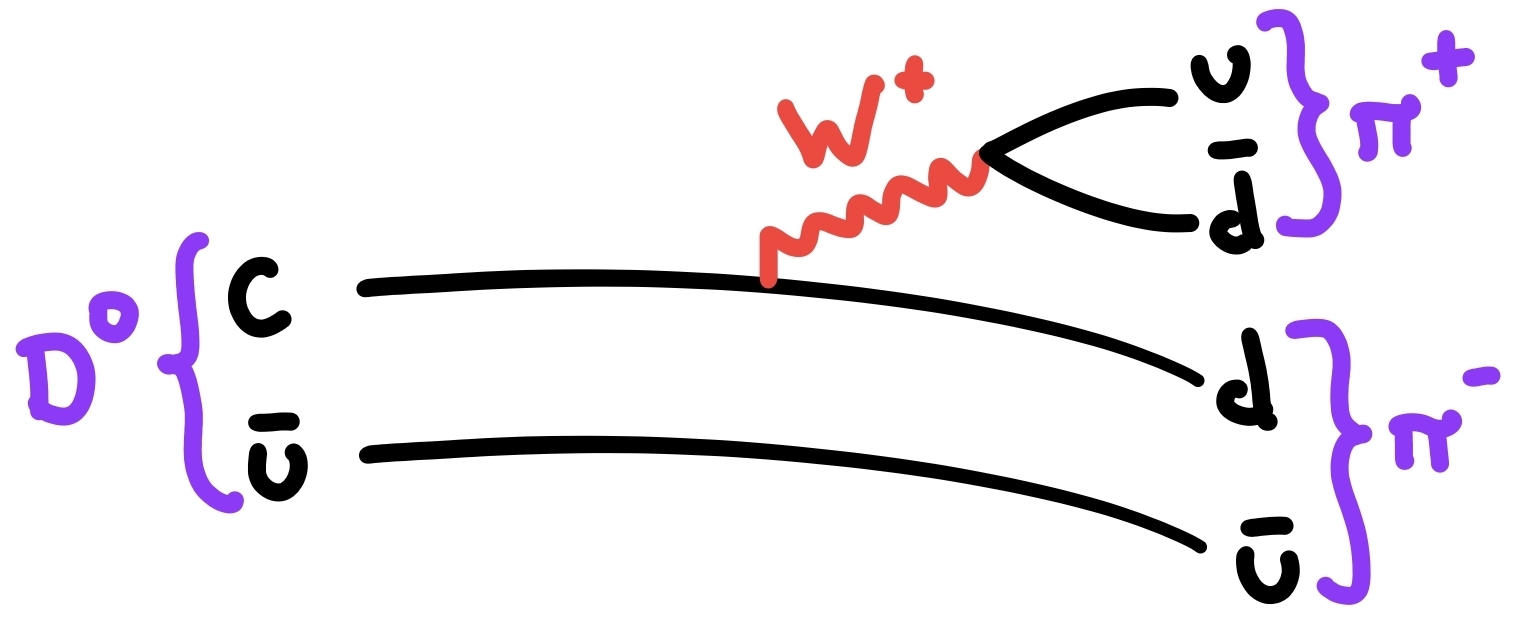} \hspace{0.5cm}
    \includegraphics[width=0.2\linewidth]{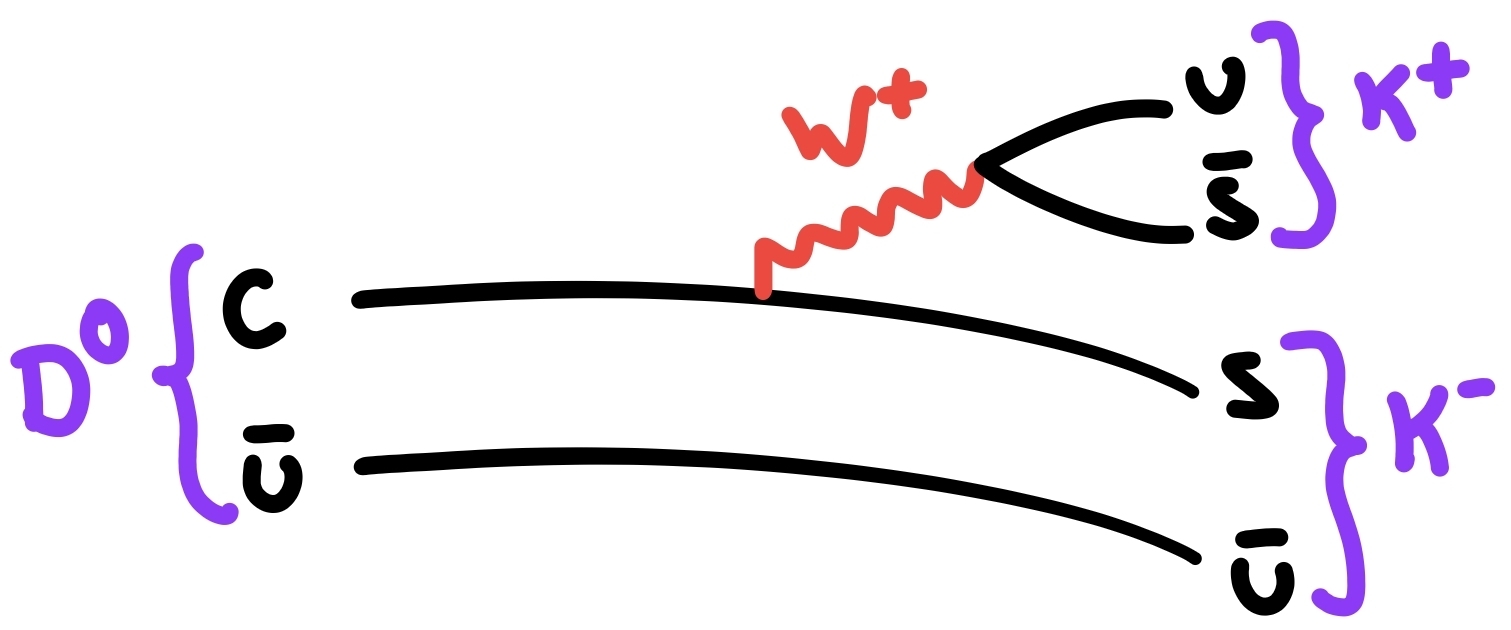} \hspace{0.5cm}
    \includegraphics[width=0.2\linewidth]{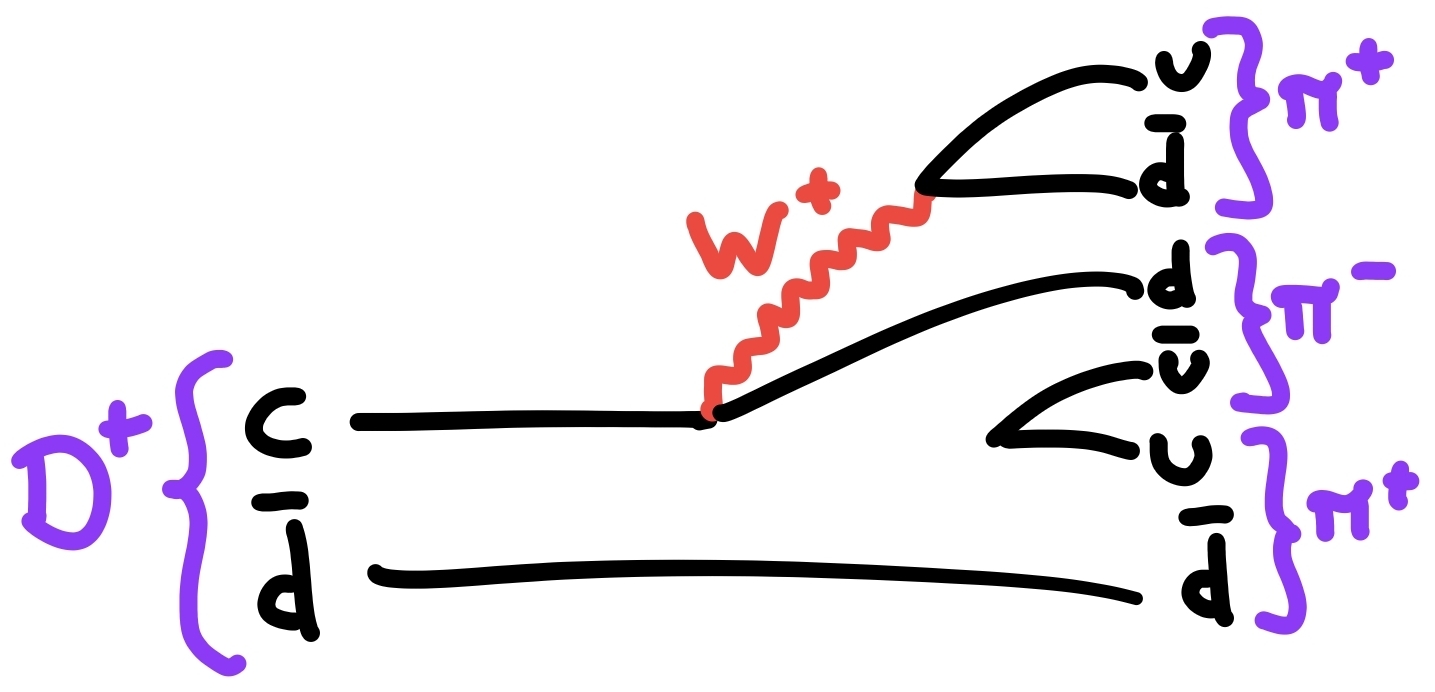} \hspace{0.5cm}
    \includegraphics[width=0.2\linewidth]{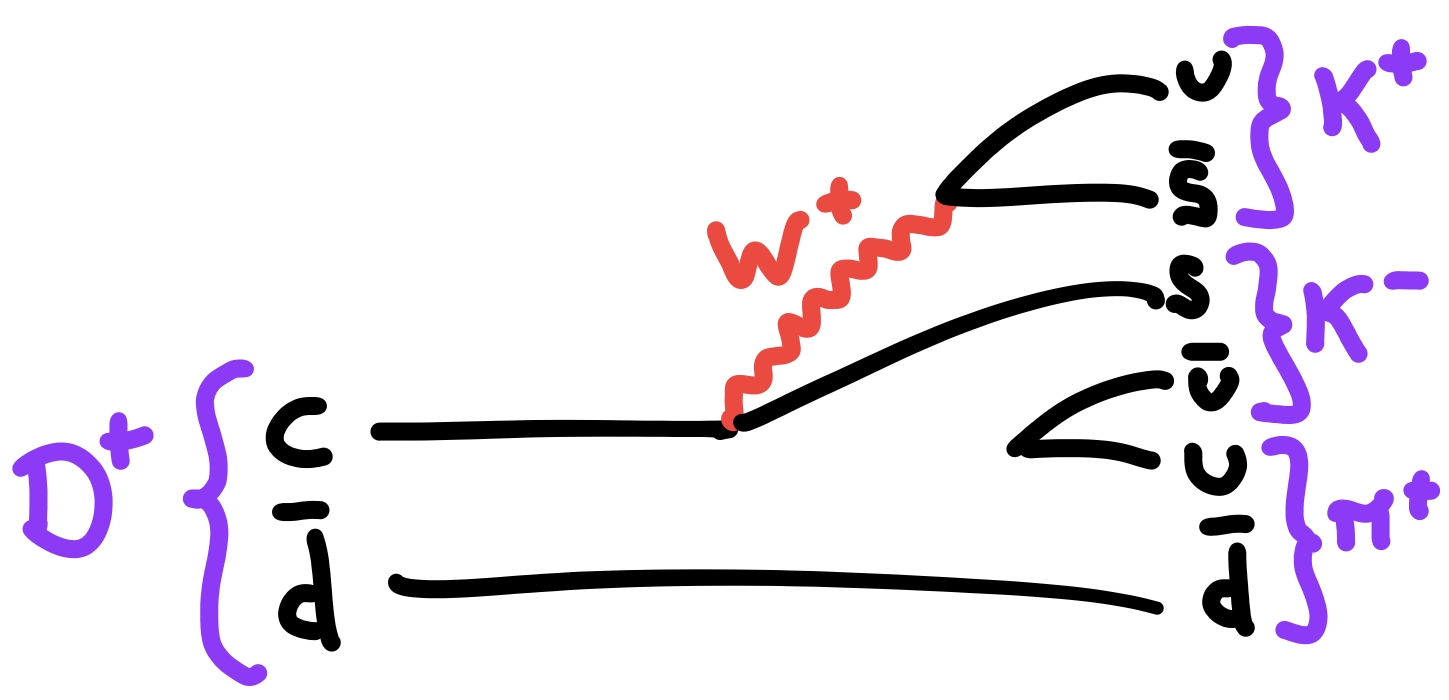}
    \caption{Tree level diagrams for $\dpp$, $\dkk$, $\dppp$ and $\dkkp$}
    \label{fig:weakvertices}
\end{figure}

\vspace{-0.3cm}
The first observation of CPV in decays of the D meson was from a combined measurement of the CP asymmetry in the $\dkk$ and $\dpp$ systems~\cite{LHCb_charmCPV}.
Thus, CPV is expected in $\dkkp$, because the weak vertex is the same as in $\dkk$, just as $\dppp$ has the same vertex as $\dpp$, as seen in Figure \ref{fig:weakvertices}. 

However, CPV has not yet been observed in charm three-body decays. The LHCb Experiment at CERN carried out a model-independent search for localised CPV in the phase space of the $\dkkp$ decay~\cite{Lhcb2024_DKKpi_Felipe}, but the results were consistent with absence of global and local CPV. 
A full amplitude analysis might be more sensitive to localized CPV in intermediate amplitudes, even in the case of a low integrated asymmetry\cite{PelaezPatBhhh}.

A novelty of this amplitude analysis is the robust inclusion of hadronic Final-State Interactions (FSI) from unitarized amplitudes and three-body effects in the amplitude description. The FSI contribute significantly to the dynamics of hadronic charm decays, and their strong phase interference with the weak phases from the CKM matrix can enhance the CP asymmetries~\cite{TIP_enhanced_charmAcp}. 

The last amplitude analysis of the $\dkkp$ decay is from 2008, carried out by the CLEO-c collaboration~\cite{Cleo2008_Dkkp}, with a sample of $\sim20000$ events and the Isobar Model to describe the data. In contrast to CLEO-c, the LHCb Run 2 data sample used in~\cite{Lhcb2024_DKKpi_Felipe} has $\mathcal{O}(10^8)$ events, offering a unique opportunity for this study.

\section{Dalitz Plot Analysis}
\label{sec:ampan}
\vspace{-0.3cm}
The dynamics of three-body decays can be fully described as a function of two of the invariant masses of pairs of the final state particles, e.g, $s_{12}=(p_1 + p_2)^2$ and $s_{13}=(p_1 + p_3)^2$, where $p_i$ is the four-momentum of a final state particle ($i=1,2,3$). The Dalitz Plot~\cite{Dalitz} then shows the decay rate as a function of the invariant masses, and relates to the total amplitude by equation \ref{eq:dalitz}.

\begin{equation}
 d\Gamma=\frac{1}{(2\pi)^3}\frac{1}{32M^3}|\mathcal{T}|^2ds_{12}ds_{13},
 \label{eq:dalitz}
\end{equation}
where $\mathcal{T}$ is the amplitude of the decay, containing all of the dynamics, and $M$ is the mass of the decaying particle. Thus, in an amplitude analysis, the model for $\mathcal{T}$ allows for a complete description of the decay dynamics. 
 
The standard approach to describe the dynamical structures in the amplitude of heavy meson decays consists of using the Isobar Model, describing it as a coherent sum of resonances,

\begin{equation}
    \mathcal{T}(s_{12}, s_{13})=c_{nr}\tau_{nr}(s_{12}, s_{13})+ \sum_k c_k\tau_k(s_{12}, s_{13}),
    \label{eq:isobar}
\end{equation}
where $k$ is a resonance label and $nr$ is the non-resonant label, $c_k, c_{nr}$ are coefficients, and $\tau_{nr}$ usually a constant. The dynamic amplitudes for the resonances, $\tau_k$, are the product of form factors, the angular dependence, and the mass-dependent lineshape, often a Breit-Wigner distribution.
However, this description lacks a Lagrangian to describe the fundamental dynamics, limiting the physical interpretation of the parameters 
\cite{MM}.

\section{Dynamic Model}
\label{sec:model}
\vspace{-0.3cm}
The dynamic model proposed here aims at modeling the decay amplitude with a natural inclusion of FSI for SU(3) meson-meson interactions with coupled channels and an isospin decomposition. To do so, it factorizes the weak vertex, separated by its different possible topologies~\cite{Chau}. These weak vertices are then dressed by including all possible two and three-body Final State Interactions. The two-body meson-meson unitarized amplitude interactions will be included using the description in \cite{MM}. The inclusion of three-body effects are left for further model refinements.

\subsection{Weak Vertex}
\begin{figure}[ht]
    \centering  \includegraphics[width=0.22\linewidth]{figures/weak_dkkp_ca.jpg} \hspace{0.02\linewidth}
    \includegraphics[width=0.22\linewidth]{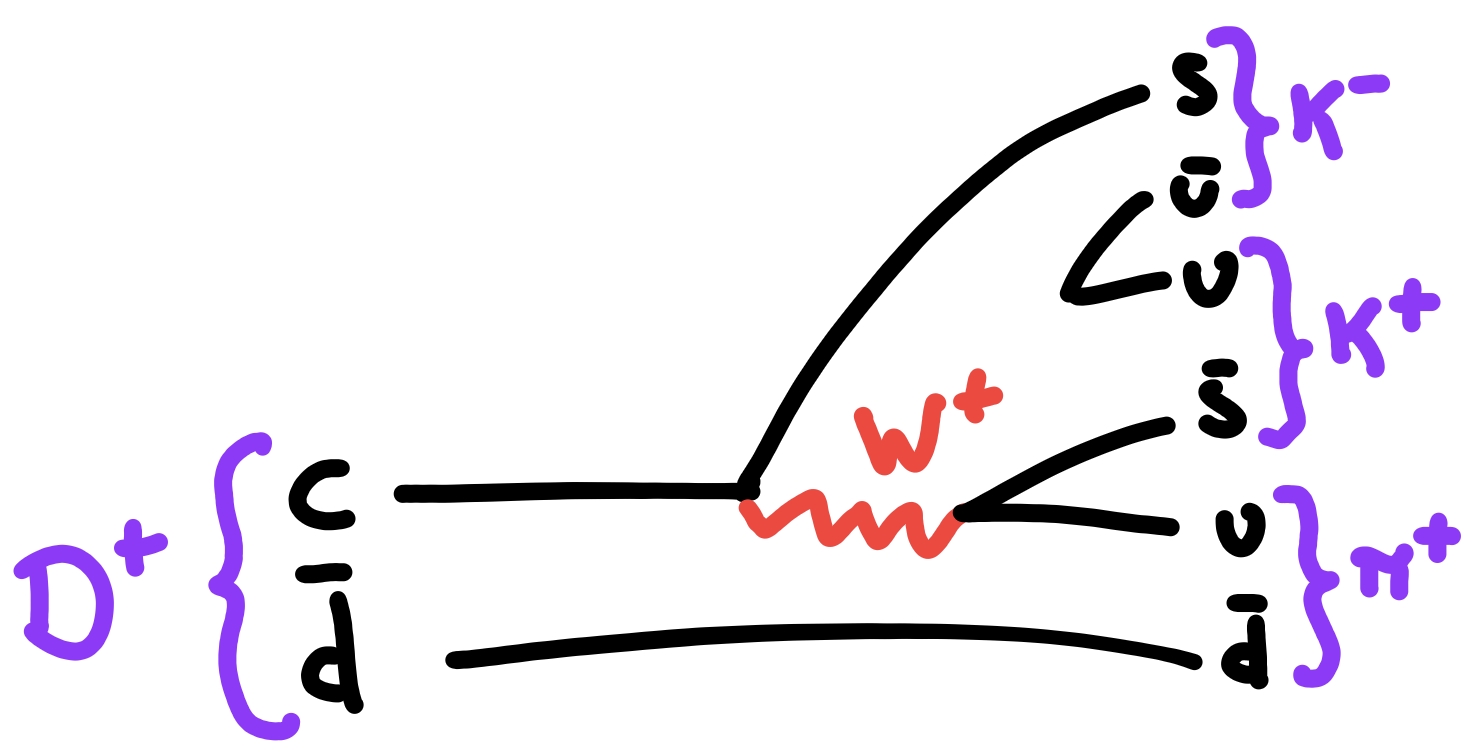} \hspace{0.02\linewidth}
    \includegraphics[width=0.22\linewidth]{figures/weak_dppp_ca.jpg} \hspace{0.02\linewidth}
    \includegraphics[width=0.22\linewidth]{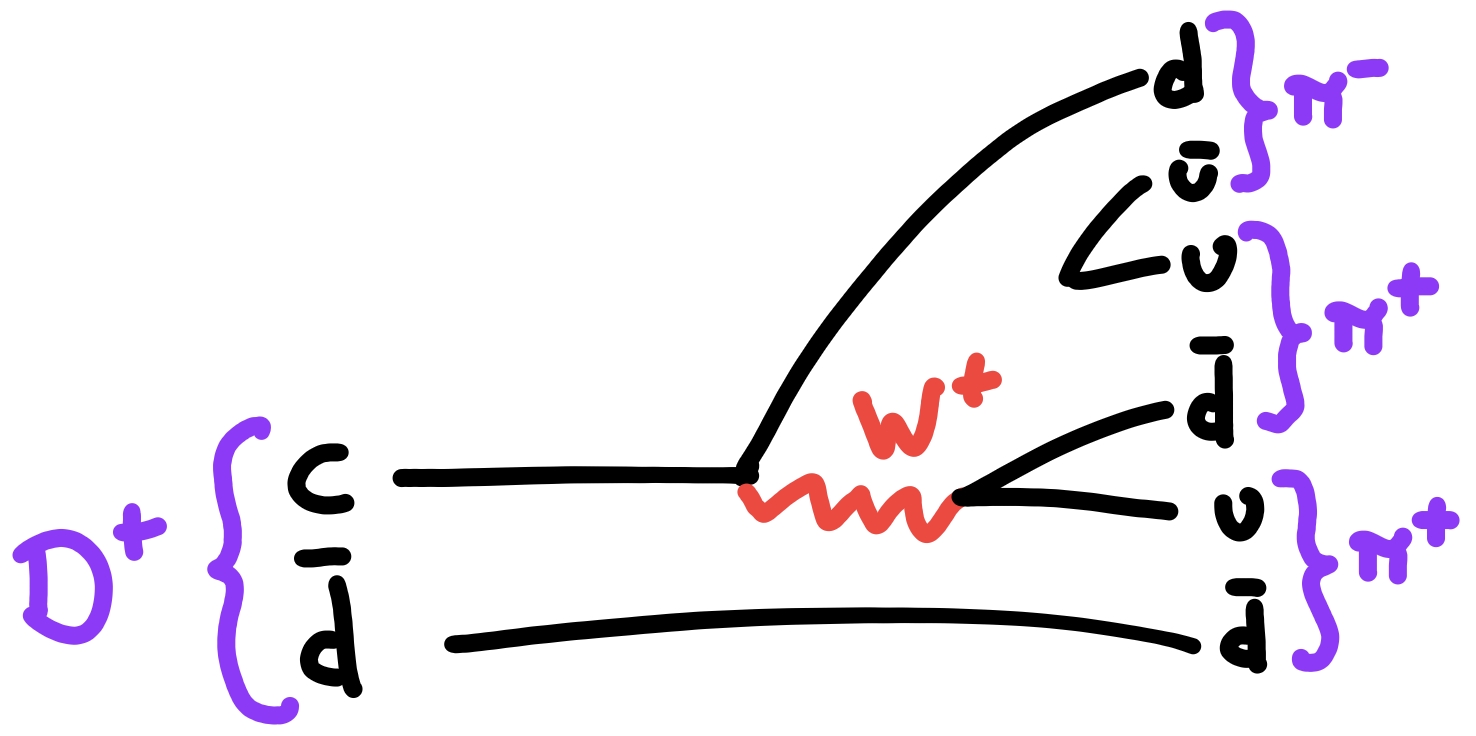}
    \caption{Weak topologies identified as leading order contributions to $\dkkp$. Diagrams are respectively: CA and CS diagrams with $V_{cs}^*V_{us}$ CKM content, CA and CS diagrams with $V_{cd}^*V_{ud}$ CKM content.}
    \label{fig:weaktopologies}
\end{figure}

\vspace{-0.2cm}
The leading order topologies are taken to be just the color allowed (CA) and color suppressed (CS) diagrams (Figure \ref{fig:weaktopologies}), as other weak topologies are not dominant. The diagrams with $V_{cd}^*V_{ud}$ Cabibbo-Kobayashi-Maskawa (CKM) matrix elements contribute to the $K^-K^+\pi^+$ final state after the introduction of $\pi\pi\to KK$ FSI.
Each weak vertex introduces a phase through its CKM matrix elements; however, the full quark-to-hadron description of the weak vertex in effective QCD theories is still an open problem. Thus, the parametrization of the weak vertices is chosen as their CKM content multiplied by a polynomial with free parameters to be fitted to data, with different parameters for vertices with different CKM elements, as in Eq. \ref{eq:weakvertex},
%
%
\begin{equation}
    \label{eq:weakvertex}
    W_q(s_{12}, s_{13})=V_{cq}^*V_{uq}\left(w_q^0 + w_q^{KK}\left(s_{12}-4m_K^2\right) + w_q^{K\pi}\left(s_{13}-\left(m_K+m_{\pi}\right)^2\right)\right).
\end{equation}
\subsection{Final State Interactions}
\vspace{-0.2cm}
The model in \cite{MM} departs from extensions of Resonant Chiral Perturbation Theory (ChPTR)~\cite{EGPR} to describe, among others, $KK$ and $K\pi$ FSI. To do so, it defines a kernel $\mathcal{K}$ for SU(3) meson-meson interactions as the sum of a contact interaction and resonance exchange(s). Higher-order diagrams can be obtained by adding interaction kernels mediated by meson loops, resulting in the expansion of FSI amplitudes in the form $\mathcal{A} = \mathcal{K}/(1-(\text{loop}\times\mathcal{K}))$.
%
This description of the expansion of FSI will be used to introduce the presence of resonances and contact interactions in the model. Their parameters, such as resonance masses and coupling constants of the effective interactions, are fitted to pseudodata generated from parametrizations of scattering amplitudes 
\cite{Pelaez2019, pelaez_pipi_2011, pelaezreview_kpi_pipikk_2022}. Those parametrizations were obtained from fits to scattering data with dispersion relation techniques.

\subsubsection{$K\pi$ and $KK$ Inputs}
\vspace{-0.1cm}
Using the description of the two-body interactions from \cite{MM}, the $K\pi\to K\pi$ amplitude can be written as in equation \ref{eq:amp012}. For the $KK$ channel, the isospin 0 ($I=0$) channel is coupled to the $\pi\pi$ channel, and the coupled amplitudes take the form of equation \ref{eq:amp00}, such that the kernels and their parameters are shared between $\pi\pi\to\pi\pi$, $\pi\pi\to KK$ and $KK\to KK$ amplitudes. But there is no $KK\to KK$ scattering data available to constrain a parametrization. Therefore, isospin 1 parameters will be obtained directly from fits to the three-body decay, but for isospin 0 the $\pi\pi\to\pi\pi$ amplitude is well known, so its parametrization was used to fix $I=0$ parameters.
{\begingroup
\renewcommand*{\arraystretch}{1.8}
\footnotesize
\begin{equation}
    \label{eq:amp012}
\mathcal{M}^{(0,1/2)}=-
    \begin{pmatrix}
        \mathcal{K}^{(0,1/2)}_{(\pi K|\pi K)} \Omega_{\pi K}^S & \mathcal{K}^{(0,1/2)}_{(\pi K|K\eta')} \Omega_{K\eta'}^S \\
        \mathcal{K}^{(0,1/2)}_{(\pi K|K\eta')} \Omega_{\pi K}^S & \mathcal{K}^{(0,1/2)}_{(K\eta'|K\eta')} \Omega_{K\eta'}^S
    \end{pmatrix}, \hspace{0.6cm}
    A^{(0,1/2)}_{(\pi K|\pi K)} = \frac{(1-\mathcal{M}^{(0,1/2)}_{22})\mathcal{K}^{(0,1/2)}_{(\pi K|\pi K)} + \mathcal{M}_{12}\mathcal{K}^{(0,1/2)}_{(\pi K|K\eta')}}{\det(\mathbbm{1}-\mathcal{M})}
\end{equation}
\endgroup}
{\begingroup
\renewcommand*{\arraystretch}{1.8}
\footnotesize
\begin{equation}
    \label{eq:amp00}
\mathcal{M}^{(0,0)}=-\frac{1}{2}
    \begin{pmatrix}
        \mathcal{K}^{(0,0)}_{(\pi\pi|\pi\pi)} \Omega_{\pi\pi}^S & \mathcal{K}^{(0,0)}_{(\pi\pi|KK)} \Omega_{KK}^S \\
        \mathcal{K}^{(0,0)}_{(\pi\pi|KK)} \Omega_{\pi\pi}^S & \mathcal{K}^{(0,0)}_{(KK|KK)} \Omega_{KK}^S
    \end{pmatrix}, \hspace{0.6cm}
    A^{(0,0)}_{(i|KK)} = \frac{(\mathbbm{1}-\mathcal{M})_{2,2-i}\mathcal{K}^{(0,0)}_{(\pi\pi|KK)} + (\mathbbm{1}-\mathcal{M})_{1,2-i}\mathcal{K}^{(0,0)}_{(KK|KK)}}{\det(\mathbbm{1}-\mathcal{M})}
\end{equation}
\endgroup}

\begin{figure}[ht]
    \centering
    \begin{minipage}[c]{0.755\textwidth}
        \centering
        \includegraphics[width=0.495\linewidth]{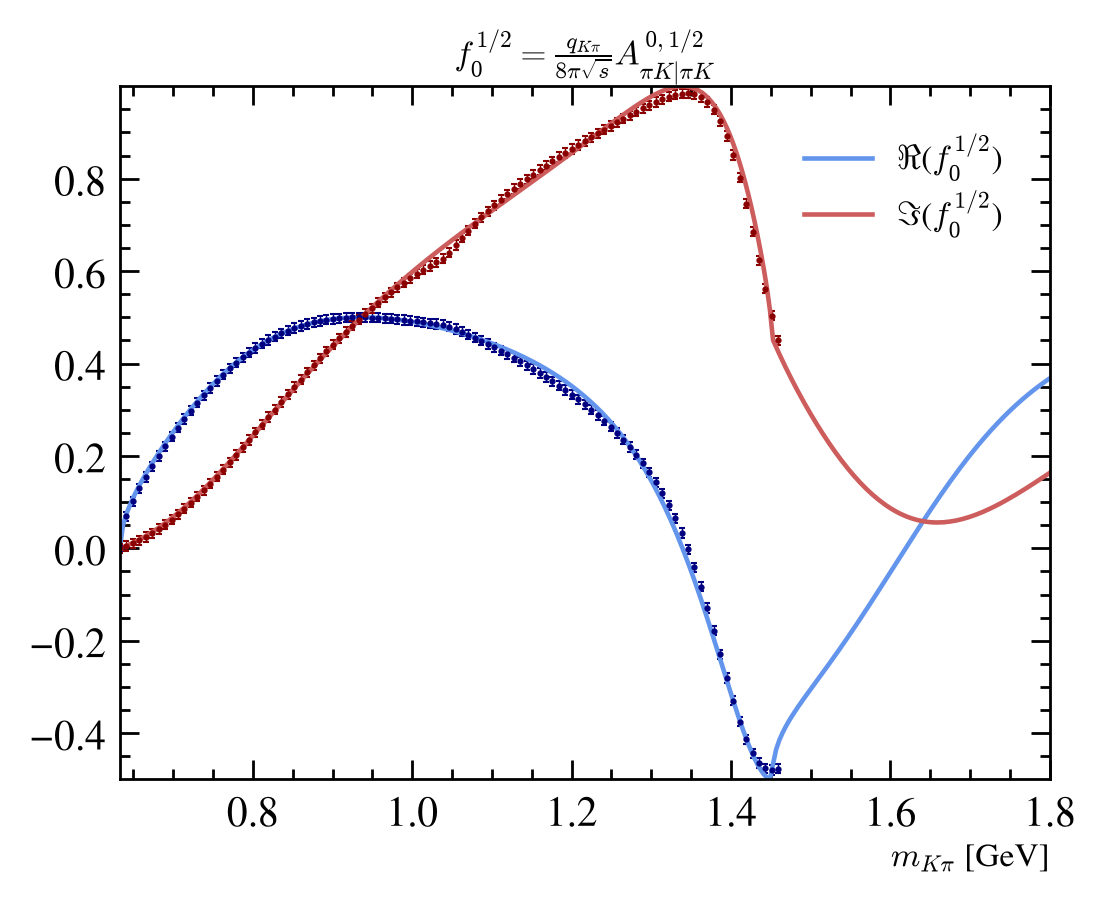}
        \includegraphics[width=0.495\linewidth]{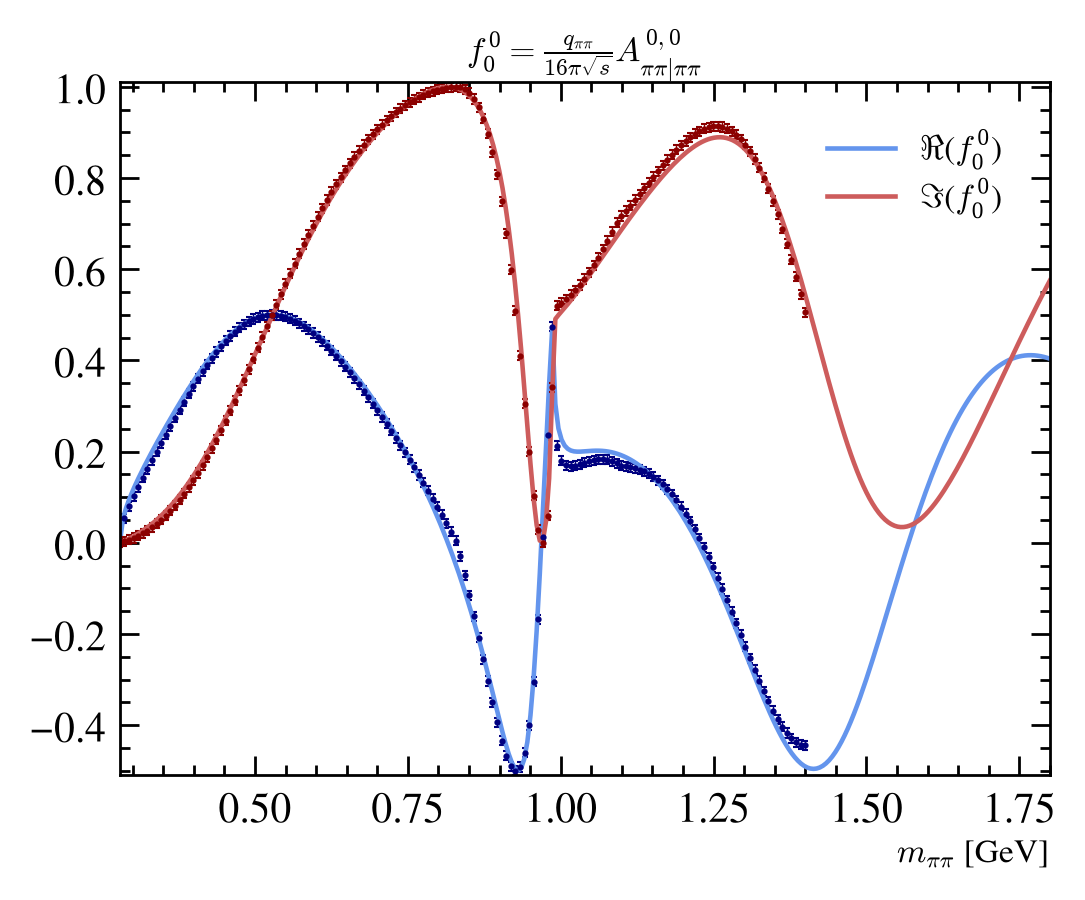}
        \captionof{figure}{Real and imaginary parts of the $\pi K\to\pi K$ and $\pi\pi\to\pi\pi$ amplitudes. Full line is the fitted model and errorbars are the generated pseudodata.}
        \label{fig:2bampfits}
    \end{minipage}
    \hfill
    \begin{minipage}[c]{0.235\textwidth}
        \centering
        \tiny
        \begin{tabular}[c]{|c|S[table-format=1.7]|}
             \hline
              $f_{\pi}$ & 0.0986(7) \\
              $f_{K}$ & 0.0904(7) \\
              $f_{\eta}$ & 0.12(1) \\
              $c_d$ & 0.0507(4) \\
              $\tilde{c}_d$ & 0.0206(3) \\
              $c_{(So|\pi\pi)}$ & 0.35(2) \\
              $c_{(So|KK)}$ & -0.062(2) \\
              $c_{(S1|\pi\pi)}$ & -0.07(1) \\
              $c_{(S1|KK)}$ & -0.0160(9) \\
              $m_{f_a}$ & 1.19690(1) \\
              $m_{f_b}$ & 0.8155(6) \\
              $\epsilon$ & -0.002(3) \\
              $c_{(K_0*|K\pi)}$ & -0.112(2) \\
              $c_{(K_0*|K\eta)}$ & 0.090(7) \\
              $m_{K_0*}$ & 1.3392(7) \\
              \hline
        \end{tabular}
        \captionof{table}{FSI amplitude parameter values ($GeV$) from simultaneous fit of $\pi K\to\pi K$ and $\pi\pi\to\pi\pi$.}
        \label{tab:fsiparams}
    \end{minipage}
\end{figure}

In equations \ref{eq:amp012} and \ref{eq:amp00}, the two-body S-wave propagator for a pair of pseudoscalars $ab$ is  $\Omega^S_{ab}$ and the kernels have the structure presented in~\cite{MM}, with free parameters such as $f_{\pi}$, $f_{K}$, $m_{K_0^*}$, $c_d$ and $c_{(K_0^*|\pi K)}$ (for $J=0, I=1/2$). The fits to the pseudodata (Figure \ref{fig:2bampfits}) yielded the parameter values in Table \ref{tab:fsiparams}, demonstrating good agreement between model and data. Once these parameters have been obtained, they can now be fixed in the fits to three-body decay data. 

\subsection{Proof of Concept: CLEO S-wave}
\begin{figure}[ht]
    \centering
    \includegraphics[width=0.39\linewidth]{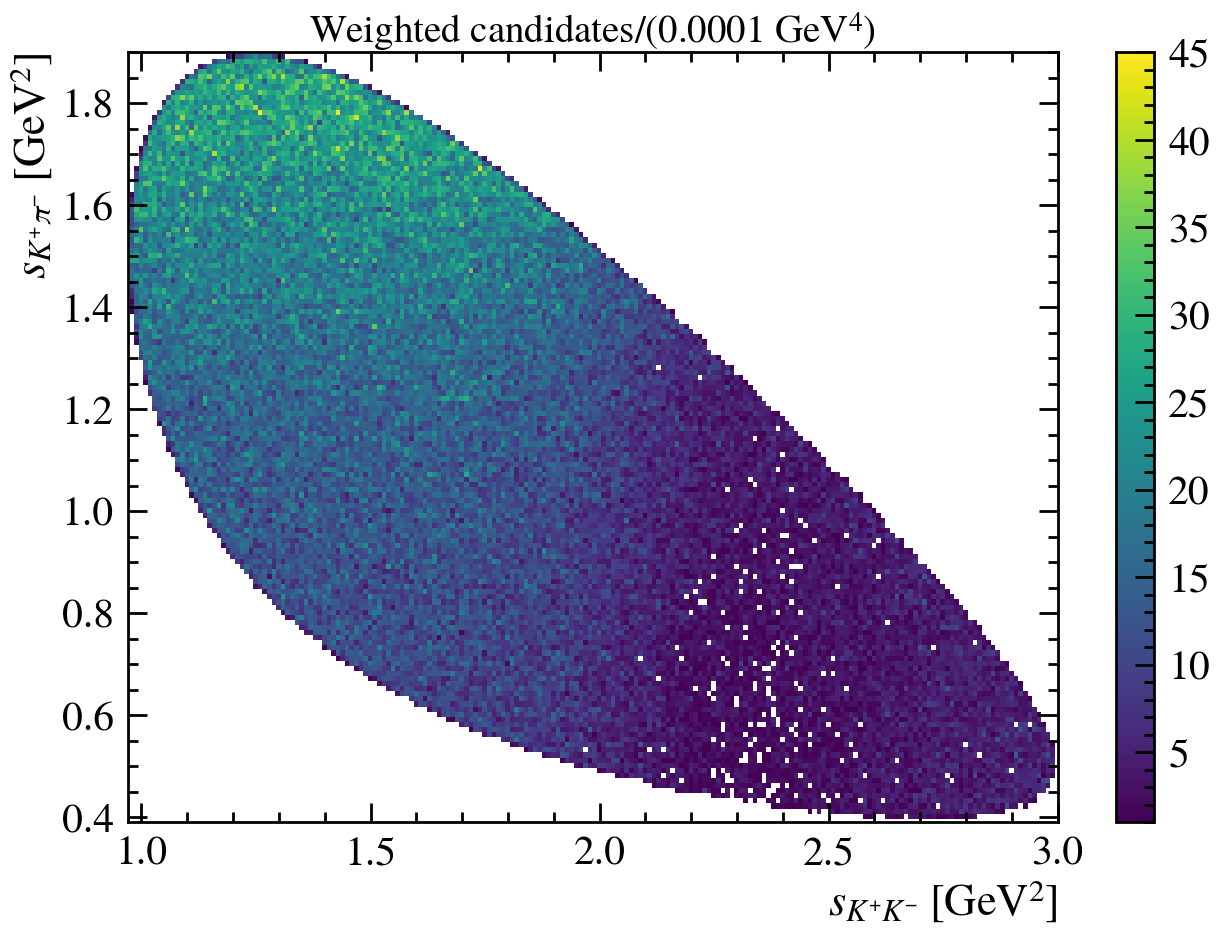}
    \includegraphics[width=0.6\linewidth]{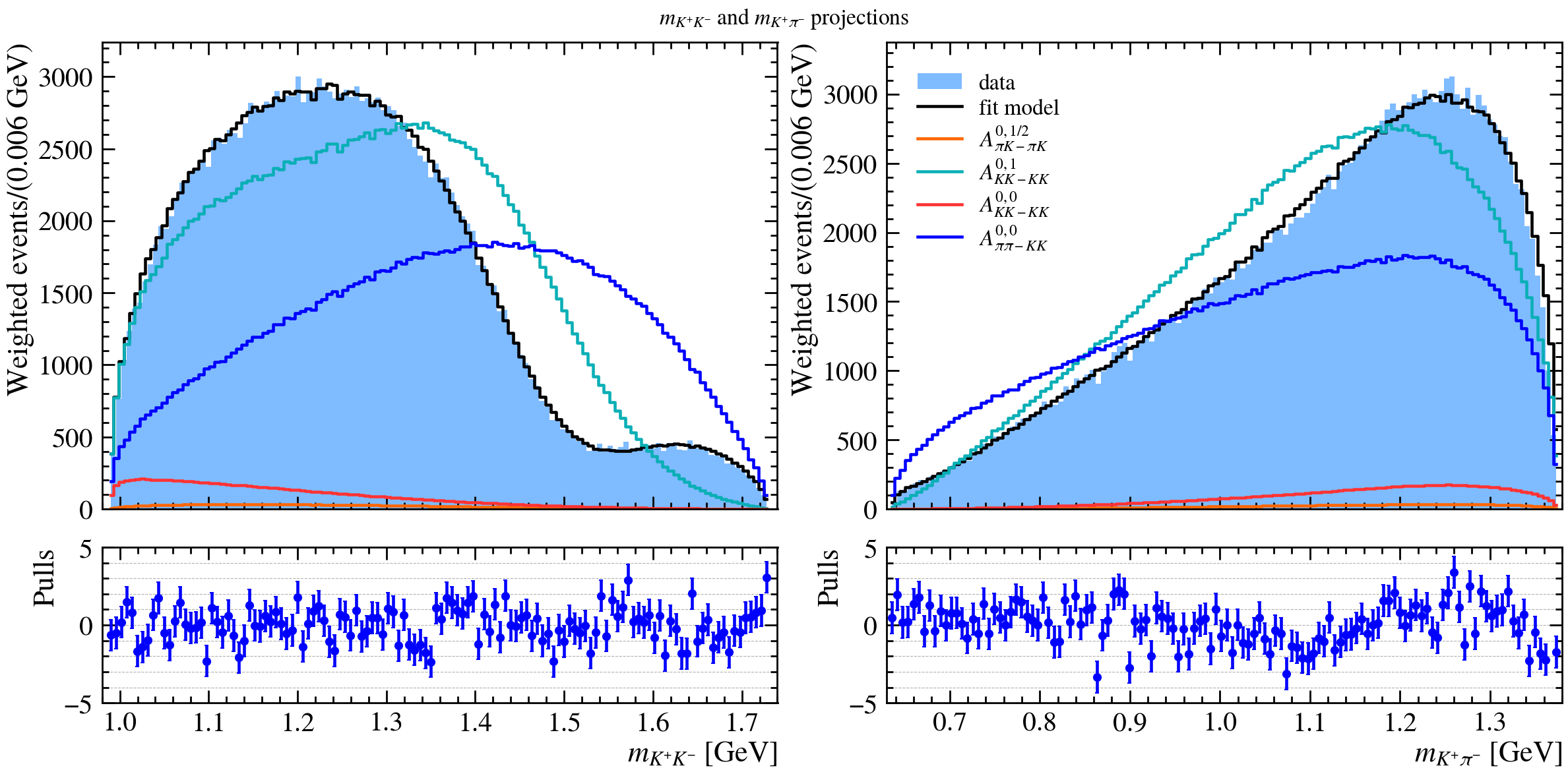}
    \captionof{figure}{Dalitz plot and fitted projections of the toy data sample.}
    \label{fig:cleofit}
\end{figure}

\vspace{-0.2cm}
With the definition of the parametrization for the weak vertex and having constrained the FSI amplitudes from fits to scattering data, the dynamic model can be used to describe the three-body decay amplitude. As a proof of concept, this is shown for the S-wave of the $\dkkp$ amplitude extracted from~\cite{Cleo2008_Dkkp}, 
which included the $a_0(1450)$ and $K_0^*(1430)$  as Breit-Wigners, and a non-resonant contribution.  A toy data sample was generated and fitted with our model (Fig.  \ref{fig:cleofit}).



The S-wave model, Eq. \ref{eq:swave}, includes $KK\to KK$ ($I=0,1$), $\pi\pi\to KK$ ($I=0$) and $K\pi\to K\pi$ ($I=1/2$) scattering amplitudes. Since there is no $KK\to KK$ scattering data, some parameters from FSI expressions were free in the fit, and the results are presented in Table \ref{tab:swaveparams}.
\begin{equation}\mathcal{T}^S=W_s\left(x^{(0,1/2)}A^{(0,1/2)}_{(\pi K|\pi K)}+x^{(0, 0)}A^{(0, 0)}_{(KK|KK)}+A^{(0, 1)}_{(KK|KK)}\right)+W_d\left(A^{(0, 0)}_{(\pi \pi|K K)}\right)
    \label{eq:swave}
\end{equation}
\vspace{-0.5cm}
\begin{table}[ht]
    \centering
    \renewcommand*{\arraystretch}{1.6}
    \setlength{\tabcolsep}{5pt}
    \tiny
    \begin{tabular}[c]{|c|c|c|c|c|c|c|c|c|c|c|c|}
        \hline
        $w_s^{0}$ & $w_s^{\pi K}$ & $w_s^{KK}$ & $w_d^{0}$ & $w_d^{\pi K}$ & $w_d^{KK}$ & $x^{(0, 0)}$ & $x^{(0, 1/2)}$ & $c_{(S1|\pi\pi)}$ & $c_{(S1|KK)}$ & $c_{(a_0|KK)}$ & $m_{a_0}$ \\
        \hline
        -1.09(4) & -0.295(14) & 0.461(21) & 0.55(9) & -1.25(6) & -1.77(6) & 0.37(7) & -0.198(24) & 0.00634(34) & 0.286(18) & -0.0789(12) & 1.4759(21) \\
        \hline
    \end{tabular}
    \caption{Parameter values from the fit to the toy sample.}
    \label{tab:swaveparams}
\end{table}

It can be seen that this model is capable of adequately describing the S-wave of this toy sample, obtained from the most accurate description of this decay to date.

\section{Conclusion}
\label{sec:conclusion}
\vspace{-0.3cm}
The amplitude model for $\dkkp$ decay under development includes a proper description of low-energy meson-meson dynamics from the extension of Resonant Chiral Perturbation Theory and parametrizes the weak vertex by unknown polynomial functions considering all the allowed leading topologies. This description overcomes some limitations of the standard Isobar Model by allowing a more complete description of Final State Interactions, as well as allowing for the inclusion of three-body effects.
By using scattering data to constrain the FSI amplitudes to their known forms, the parametrization of the weak vertex can be obtained with more precision, and a proof-of-concept study showed that this model is adequate to describe, at least, the S-wave of the $\dkkp$ decay as proposed by the CLEO collaboration. Thus, this model is expected to improve the phenomenological description of the amplitude and yield a better result in the amplitude analysis of the $\dkkp$ decay. Given this better precision in describing the decay from a theoretical perspective, the amplitude analysis could indicate signs of CPV in the dynamics of $\dkkp$. 

\acknowledgments
\vspace{-0.3cm}
This work was financed, in part, by the São Paulo Research Foundation (FAPESP), Brasil. Process Numbers 2024/15576-0 and 2025/21317-0.

{
\small
\singlespacing
\bibliographystyle{JHEP}
\bibliography{references}
}

\end{document}